\documentclass[12pt]{iopart}
\usepackage[utf8]{inputenc}
\usepackage[american,british]{babel}
\usepackage[T1]{fontenc}
\usepackage[pdftex]{graphicx}  
\usepackage{xcolor}
\usepackage{dcolumn}
\usepackage{bm}
\usepackage{verbatim}
\usepackage{ulem}
\usepackage{epsfig}
\usepackage{color}
\usepackage{xfrac}
\usepackage{hyperref}
\usepackage{iopams} 
\usepackage{setstack}

\newcommand{\bra}[1]{\left\langle #1 \right|}
\newcommand{\ket}[1]{\left| #1 \right\rangle}

\renewcommand{\Im}{\mathfrak{Im}}

\begin{document}

\title{Scattering of mesons in quantum simulators}

\author{Federica Maria Surace$^{1,2}$\footnote{fsurace@sissa.it}, Alessio Lerose$^3$\footnote{alessio.lerose@unige.ch}}
\address{$^1$ SISSA --- International School of Advanced Studies, via
Bonomea 265, 34136 Trieste, Italy}
\address{$^2$ The Abdus Salam International Center for Theoretical Physics, Strada Costiera 11, 34151 Trieste, Italy}
\address{$^3$ Department of Theoretical Physics,
University of Geneva, Quai Ernest-Ansermet 30,
1205 Geneva, Switzerland}

\begin{abstract}
Simulating real-time evolution in theories of fundamental interactions represents  one of the central challenges in contemporary theoretical physics. 
Cold-atom platforms
stand as 
promising candidates to realize quantum simulations of 
non-perturbative phenomena in gauge theories, such as vacuum decay and {hadron} collisions, in prohibitive conditions  for direct experiments. 
In this work, we demonstrate that present-day quantum simulators can {imitate}
linear particle accelerators, giving access to S-matrix measurements of elastic and inelastic meson collisions in {low-dimensional} Abelian gauge theories.
Considering for definiteness a $(1+1)$-dimensional $\mathbb{Z}_2$-lattice gauge theory realizable with 
{Rydberg-atom} arrays, we present
protocols to observe and measure selected meson-meson scattering processes.
We provide a benchmark theoretical study of scattering amplitudes in the regime of large fermion mass, including an exact solution valid for arbitrary 
coupling strength.
This allows us to discuss the occurrence of inelastic scattering channels, featuring the production of new mesons with different internal structures.
We present numerical simulations of {realistic} wavepacket collisions, which reproduce the predicted cross section peaks. This work highlights the potential of quantum simulations to give unprecedented access to  real-time  scattering dynamics.
\end{abstract}

\maketitle

\section{Introduction}
While implementing {fault-tolerant} quantum computations still requires significant technological advances, highly controllable quantum devices with hundreds of qubits are already being realized in various experimental platforms
~\cite{Nori2014}. 
{The possibility of accessing real-time dynamics and strongly correlated quantum many-body states
opens numerous avenues in the theoretical research. 
One of the most promising 
directions is represented by the quantum simulation of high-energy physics phenomena~\cite{Wiese:2013kk,ZoharReview,DalmonteMontangeroReview,banuls2020review,KarzeevDigital}. 
In the last decade, substantial efforts have been devoted to the implementation of gauge-invariant Hamiltonian dynamics \cite{Zohar2012,Tagliacozzo:2012kq,BanerjeeQLMmixtureProposal,Banerjee2013,Egusquiza2015,kasper2017implementing,davoudi2020search,YangHaukeNature20}.
Recently, certain aspects of the physics of vacuum decay have been explored with trapped ions~\cite{ExpPaper} and Rydberg atom arrays~\cite{BernienRydberg,SuraceRydberg}.}
A challenging problem in high-energy physics is simulating 
{collisions of complex composite particles.}
In quantum chromodynamics (QCD), a first-principle estimation of 
the distribution of particles produced by hadron scattering would facilitate the search for new physics beyond the Standard Model; {moreover, heavy-ion collisions provide fundamental information on the deconfinement transition and on the early Universe evolution~\cite{RajagopalWilczek}}.
Although simulating 
higher-dimensional non-Abelian gauge theories is still a far-fetched goal, it is of great interest to understand whether quantum simulators are already capable of studying the scattering of composite particles in a strong coupling regime, 
{at least in simplified  settings. Lower-dimensional gauge theories~\cite{SchwingerModel,tHooftModel} exhibit a  tractable version of particle confinement leading to an analog of quark-antiquark bound states (\textit{mesons}). Real-time dynamics of variants of these models witnessed recent developments in both classical ~\cite{PichlerMontangeroQLMTensorNetwork,Buyens:2014cs} and quantum~\cite{SuraceRydberg,ExpPaper,YangHaukeNature20,TanConfinementExperiment} simulations, opening the door to investigations of the simplest instances of collisions between complex structured objects arising from confinement. }

\begin{figure}[t]
    \centering
    \includegraphics[width=0.7\textwidth]{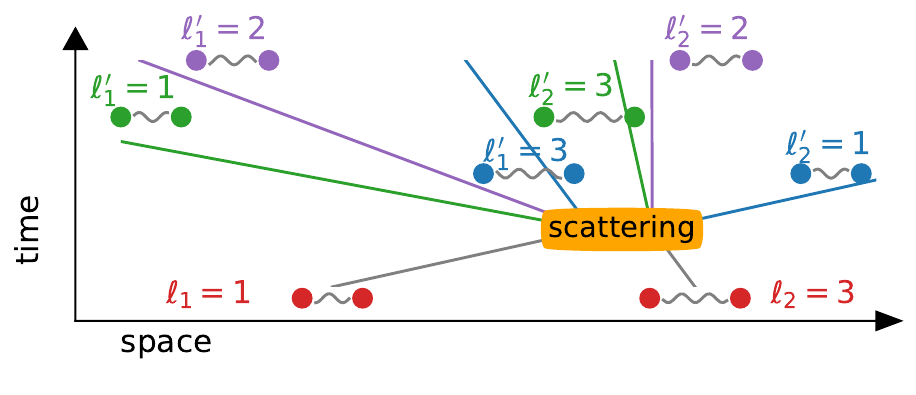}
    \caption{
    Sketch of a
    scattering event:  {The collision of two incoming mesons with internal quantum numbers $\ell_1$, $\ell_2$ generates a superposition of several possible outcomes, labelled by the quantum numbers of outgoing mesons}.
    }
    \label{fig:cartoon}
\end{figure}

In this work, we demonstrate that present-day quantum simulators allow to investigate selected meson collisions in {$1+1$-dimensional}
Abelian lattice gauge theories (LGTs), as sketched in Fig.~\ref{fig:cartoon}, {mimicking} scattering experiments with particle accelerators. 
{Quantum simulators offer unprecedented access to full real-time resolution of a complex collision event and to the quantum correlations thereby generated.}
Here, we particularly focus on 
the production of new mesonic species, i.e., \textit{inelastic} events redistributing internal and kinetic energies of mesons emerging from the collision.
We propose protocols to experimentally observe this with current facilities, and provide a benchmark theoretical study of scattering amplitudes.
While we consider a controlled regime where exact numerical simulations can be pushed and compared with analytical results, 
 quantum simulators may explore conditions inaccessible to traditional methods, including the continuum limit of quantum field theories.

The paper is organized as follows. 
In Sec.~\ref{sec_model}, we introduce the $\mathbb{Z}_2$-LGT analyzed throughout, and discuss particle confinement and the resulting mesonic spectra and wavefunctions in the regime of large particle mass.
In Sec.~\ref{sec_scattering}, we give a theoretical study of meson-meson scattering amplitudes, based on an exact solution of the Schr\"{o}dinger equation. 
We discuss elastic and inelastic processes, and benchmark the results against numerical simulations.
Finally, in Sec.~\ref{sec_quantumsim}, we propose concrete protocols to prepare, simulate and observe meson scattering with present-day quantum simulators (e.g., Rydberg-atom arrays). 
The appendices contain various additional details on the discussion and computations in the main text.
In \ref{sec_gauge} we report additional details on gauge invariance and confinement in the model under consideration in the main text. 
In \ref{sec_mapping} we prove the exact mapping of its dynamics in the gauge-neutral sector onto those of the quantum Ising chain in a tilted magnetic field.
In \ref{sec_2body} and \ref{sec_4body} we provide more details on the exact solution of the two- and four-fermion problem, i.e., on the spectra of mesons and their scattering amplitudes, in the limit of large fermion mass.
In \ref{sec_current} we derive the analytic expression of the meson current, we discuss its physical meaning and we prove the associated continuity equation.
Finally, in \ref{sec_finitemass} we summarize and discuss the effects of having a finite fermion mass.

\section{Confinement and mesons}
\label{sec_model}
{Particle confinement is a non-perturbative phenomenon arising in certain gauge theories, 
which {consists} in the absence of \textit{charged} asymptotic states: all stable excitations of the theory above the ground state are \textit{neutral} bound states of fermionic charges \cite{Weinbergbook2}.}
{In the context of QCD, confinement underlies the fact that quarks can only be observed in composite structures such as mesons and baryons. 
Despite the fundamental difference between particle confinement in QCD in $(3+1)$ dimensions and in lower-dimensional models~\cite{SchwingerModel,tHooftModel}, the emergent composite particles share some basic properties, making the latter convenient settings to gain insights into difficult aspects of the theory. 
In this work we will be concerned with $(1+1)$-dimensional LGTs of this kind.
}

For the sake of definiteness, we will focus on the $\mathbb{Z}_2$-LGT defined by the following Hamiltonian~\cite{LeroseSuraceQuasilocalization,borla2020gauging}:
\begin{equation}
\label{eq_HZ2}
    H= m \sum_j c^\dagger_{j} c_j \, +\frac{\tau}{2}  \sum_j  \sigma^z_{j+1/2}
    +w \sum_j 
\big(c_{j}^\dagger-c_j\big) \, \sigma^x_{j+1/2} \,
\big(c_{j+1} + c_{j+1}^{\dagger} \big).
\end{equation}
In this equation, $c_{j}^\dagger$ and $c_j$ denote creation and annihilation operators of spinless fermions of mass $m>0$ on the sites $j \in \mathbb{Z}$ of a one-dimensional lattice, and $\sigma^{x,y,z}_{b}$ denote \mbox{spin-$1/2$} operators (Pauli matrices) acting on the bonds $b \in \mathbb{Z}+1/2$ of the lattice, representing a gauge field with string tension $\tau$.
%
Interactions, with coupling strength $w$, are such that all the local operators
$
G_j=\sigma_{j-1/2}^z \, \sigma_{j+1/2}^z (1-2c^\dagger_{j}c_j)
$
are conserved, i.e., $[G_j,H]=0$. These operators satisfy $G_j^2 \equiv 1$ and thus generate local $\mathbb{Z}_2$ symmetries. 
Here we focus on the neutral gauge sector, 
i.e., the subspace characterized by $G_j|\psi\rangle = |\psi\rangle$ for all $j$.
(For more details, see~\ref{sec_gauge}.)

The LGT in Eq.~(\ref{eq_HZ2}) exhibits particle confinement for
$m > 2|w|$, $\tau \neq 0$.
%
{By }gauge invariance, 
a string of excited gauge field 
extends between 
two charges created out of the vacuum, 
{inducing} a confining potential $V(r)\propto r$ that grows unbounded at large distances $r$. 
Thus, the 
{excitations} form a discrete tower of \textit{neutral}~\footnote{Note that the $\mathbb{Z}_2$-charge is defined modulo $2$, i.e., particle and antiparticle are the same object, so a two-fermion bound state is neutral.} bound states {(termed \textit{mesons}, in analogy with QCD)}, labelled by their internal quantum number $\ell=1,2,\dots$ and their center-of-mass momentum $k$.

In the large-$m$ limit, mesonic spectra $\mathcal{E}_\ell(k)$ and wavefunctions $\psi_{\ell,k}(j_1,j_2)$ can be determined exactly by solving the reduced two-body problem, governed by the projection of $H$ in Eq.~(\ref{eq_HZ2}) onto the two-fermion sector spanned by states $\{\ket{j_1<j_2}\}$ (labelled by the positions of the two fermions along the chain).
For definiteness, we assume $\tau>0$ from now on.
The projected Hamiltonian $H_{\text{2-body}}$ consists of nearest-neighbor hopping terms of amplitude $w$ for the two particles, plus the diagonal confining potential $\tau \, (j_2-j_1)$.
The problem is solved by switching to the center-of-mass and relative variables, $s=j_1+j_2$, $r=j_2-j_1>0$, and using the ansatz
$
    \psi(s,r) = e^{iks} \phi_k(r)
$.
The 
exact solution \cite{FogedbyTwoKinkSolution,LeroseSuraceQuasilocalization} yields the quantized mesonic spectra
\begin{equation}
\label{eq_mesonspectrum}
    \mathcal{E}_\ell(k)= \tau \, \nu_\ell(2\tilde{w}_k/\tau) 
\end{equation}
with \mbox{$\ell=1,2,\dots$, $\tilde{w}_k=2w\cos k$, $k\in[-\pi/2,\pi/2)$},  and $\nu_\ell(x)$ is the $\ell$-th (real) 
zero of the map
$a \mapsto \mathcal{J}_a(x)$, where $\mathcal J_a(x)$ is the Bessel function~\footnote{The momenta $k$ and $k+\pi$ generate the same solution up to a phase: Since $\mathcal{J}_{\alpha}(-z)=e^{i\pi \alpha}\mathcal{J}_{\alpha}(z)$, when $k\mapsto k+\pi$ the wavefunction $\Psi$ gets multiplied by $(-)^s e^{i\pi(r-\nu_n)}=e^{-i \pi \nu_n} (-)^{2 j_2}=e^{-i \pi \nu_n}$, i.e., a global phase.}.
The associated mesonic wavefunctions read
\begin{equation}
\label{eq_mesonwavefunction}
    \psi_{\ell,k}(s,r) = e^{iks} \mathcal{J}_{r-\nu_\ell(2\tilde{w}_k/\tau)}(2\tilde{w}_k/\tau).
\end{equation}
{As an example, Fig.~\ref{fig:momentumdistribution}-(a)} below reports a plot of the lowest mesonic spectra $\ell=1,2,3$ {for  $w/\tau=0.6$}.
For the derivation of these results, see  \ref{sec_2body}.

{While suppressing dynamical fluctuations of the fermion number, the large-$m$ limit encompasses the regime $|w| \gg |\tau|$, where mesons are dominated by strong quantum fluctuations in their spatial extent.
A large but finite fermion mass $m$ only produces a perturbative dressing of the vacuum and of mesons, which can be explicitly computed order by order via the so-called Schrieffer-Wolff transformation ~\mbox{\cite{FrohlichSchriefferWolff,LossSchriefferWolff,McDonaldSchiefferWolffHubbard}}. 
For instance, the first correction involves next-nearest-neighbor fermion hopping with amplitude $w^2/2m$. 
Using this approach, the \mbox{large-$m$} analysis of meson dynamics can be systematically modified to achieve the desired accuracy for large but finite $m$, limited only by the practical complexity of high-order computations.
Thus, for simplicity, here we will focus on the limit of large fermion mass;
for more details on finite-$m$ effects, see \ref{sec_finitemass}}.


\section{Scattering amplitudes}
\label{sec_scattering}
{We first provide a theoretical analysis of meson-meson scattering. 
We present an exact solution of the problem in the regime of large fermion mass $m$. The predictions of elastic and inelastic cross section peaks, together with our numerical simulations, provide a non-trivial benchmark for quantum simulations.} {While our solution is valid for arbitrary coupling strength $w/\tau$ and arbitrary incoming states, quantum simulations turn out to be easiest for $w /\tau \approx 1$ and low mesonic quantum numbers, as discussed below.}

\begin{figure}[t!]
    \centering
    \includegraphics[width=0.7\textwidth]{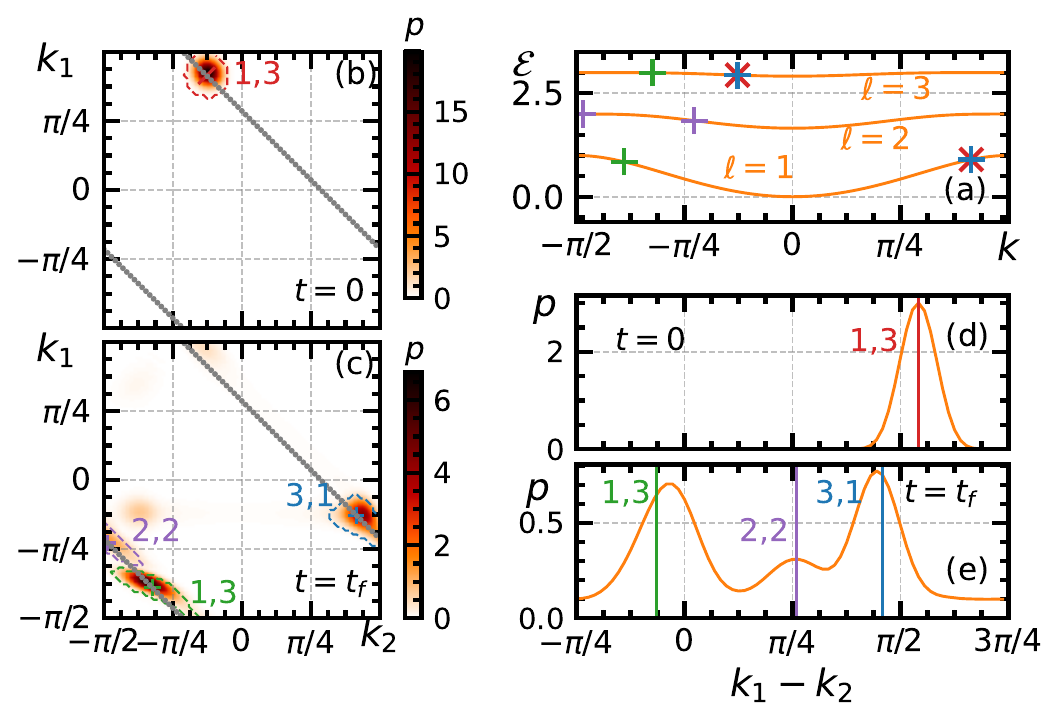}
    \caption{
    Mesonic wavepackets collision.
    (a) Spectra $\mathcal{E}_\ell(k)$ of the lightest mesons for the \mbox{$\mathbb{Z}_2$-LGT} in Eq.~(\ref{eq_HZ2}) with $\tau=1$, $w=0.6$ and $m\gg \tau$. The crosses indicate the momenta and energies of the two mesons in the incoming (red) and outgoing  (purple, blue, green) states. 
    (b-e) Probability density of the meson momenta $p(k_1, k_2)$ (b,c) and of the relative momentum $p(k_1-k_2)$ (d,e) at time $t=0$ (b,d) and $t=t_{f}=50$ (c,e).
    The dashed contours in panel~(c) mark the regions~{$p>0.25$}.
    }
    \label{fig:momentumdistribution}
\end{figure}

Armed with the mesonic spectra $\mathcal{E}_\ell(k)$, 
we consider the scattering of two incoming mesons with quantum numbers $\ell_{1,2}$, approaching each other with definite momenta $k_{1,2}$. 
The open elastic and inelastic scattering channels can be found by a kinematic analysis, which consists in determining the set of the outgoing quantum numbers $\{(\ell'_1,k'_1),(\ell'_2,k'_2)\}$ compatible with the incoming ones 
by conservation of total energy and momentum:
\begin{equation}
\label{eq_kinematic}
\left\{
    \eqalign{
        E\equiv\mathcal{E}_{\ell_1}(k_1)+\mathcal{E}_{\ell_2}(k_2) = 
        \mathcal{E}_{\ell'_1}(k'_1)+\mathcal{E}_{\ell'_2}(k'_2) \, , \cr
        K\equiv k_1 + k_2 = k'_1 + k'_2\quad  {\rm mod}\; \pi.
    }
    \right.
\end{equation}
For all choices of incoming states, there always exist two {elastic} solutions, called {\textit{transmitted} and \textit{reflected}}, having {$(\ell'_1,\ell'_2)=(\ell_2,\ell_1)$ and $(\ell_1,\ell_2)$} respectively. 
The existence of 
inelastic channels, instead, is not guaranteed for generic incoming states~%
\footnote{For intermediate ratios $\tau/w$ (a condition that best suits experiments, see below), it can be seen that inelastic channels are favoured when at least one  incoming meson is ``heavy'', i.e., $\ell_{2} > 1$. The reason is that, for sufficiently small $w/\tau$, the sum of the quantum numbers $\ell_1+\ell_2$ is conserved in the scattering. This is a consequence of the conservation of the total energy and of the fact that $\mathcal{E}(\ell,k)\simeq \tau\ell$ in this limit.  The example in Fig.~\ref{fig:momentumdistribution} comprises an inelastic channel $(\ell_1,\ell_2)=(1,3) \to (\ell'_1,\ell'_2)=(2,2)$.}.

The conservation of the number of fermions allows to derive a continuity equation, which defines an associated mesonic current, as derived in~\ref{sec_current}. 
The conservation of the total current across the collision yields a constraint on the scattering amplitudes of open channels. The fraction associated with each outgoing asymptotic solution has the physical meaning of a total cross section, as it can be identified with the probability {$P_{\ell_1',\ell_2'}$} of detecting that particular scattering outcome in the asymptotic future~\cite{norsen2009}.

Determining the scattering amplitudes and cross sections requires solving the four-fermion Schr\"{o}dinger equation.
We thus consider the  effective Hamiltonian $H_{\rm {4-body}}$ for the four-body problem, i.e., \mbox{Eq.~(\ref{eq_HZ2})} projected to the four-fermion subspace spanned by the states $\{\ket{j_1<j_2<j_3<j_4}\}$. This consists of the hopping terms of amplitude $w$ for the four particles, and the two diagonal  confining pairwise potentials $\tau(j_2-j_1)$ and $\tau(j_4-j_3)$.
We formulate the ansatz $\psi_{\ell_1, q}(s_1,r_1) \psi_{\ell_2, K-q}(s_2,r_2)$,
where $r_{1,2},s_{1,2}$ are the relative distance and the center-of-mass position for the two mesons, the single-meson wavefunctions $\psi$ are defined as in Eq.~(\ref{eq_mesonwavefunction}), but, crucially, the momentum $q \in\mathbb{C}$  
is allowed to span the complex plane. 
The ansatz above represents an admissible asymptotic solution, 
provided $q \in \mathbb{C}$ satisfies
the \textit{complex} energy condition
\begin{equation}
\label{eq_energycondition}
  \nu_{\ell_1}(2\tilde{w}_{q}/\tau)
 +
 \nu_{\ell_2}(2\tilde{w}_{K-q}/\tau)
 \; = \; E / \tau \, ,
\end{equation}
where the total energy $E$ and momentum $K$ are determined by the incoming state {$\{(\ell_1,k_1=q_{\text{in}}),(\ell_2,k_2=K-q_{\text{in}})\}$, with $q_{\text{in}}\in\mathbb{R}$, and $\nu_{\ell}(w)$ is here a complex zero of the analytic function $z\mapsto \mathcal{J}_z(w)$, labelled by $\ell \in \mathbb{N}$}.
We index by $\alpha \in \mathbb{N}$ all the triplets $(\ell_1^\alpha, \ell_2^\alpha, q_\alpha)$ which satisfy Eq.~\ref{eq_energycondition}.
For a given incoming state, 
the exact solution $\Psi$ of the scattering problem is expressed by a linear superposition including the incoming state and all compatible \textit{outgoing}  (i.e., with {outgoing} currents) and \textit{evanescent} (i.e., with $\Im (q) < 0$) asymptotic solutions:
\begin{equation}
\fl
    \Psi(s_1,r_1,s_2,r_2)= \psi_{\ell_1, k_1}(s_1,r_1) \psi_{\ell_2, k_2}(s_2,r_2)
    + \sum_{\alpha}A_{\alpha} \, \psi_{\ell_1^\alpha, q_\alpha}(s_1,r_1) \psi_{\ell_2^\alpha, K-q_\alpha}(s_2,r_2).
\end{equation}
The wavefunction $\Psi$ solves the Schr\"{o}dinger equation in the full region $j_3-j_2=(s_2-s_1-r_1-r_2)/2>0$~\footnote{The additional conditions $j_2-j_1=r_1>0$, $j_4-j_3=r_2>0$ are automatically satisfied by the ansatz.}.
Due to Pauli exclusion at $j_2=j_3$, the equation forces the boundary condition $\Psi|_{s_2-s_1-r_1-r_2=0}\equiv 0$, which determines the coefficients $A_{\alpha}$, including the scattering amplitudes of open channels. 
In fact, this condition gives rise to an infinite set of inhomogeneous linear equations on varying $r_{1,2} = 1, 2, \dots$ for the infinitely many unknowns $A_1,A_2,\dots$ 
The very nature of confinement, though, provides a natural truncation for this hierarchy: For $q_\alpha \in \mathbb{R}$, the meson wavefunctions  are bound states, and thus fall off rapidly for large distances; for complex solutions $q_\alpha \notin \mathbb{R}$, the normalizability condition \mbox{$\Im(q) < 0$} guarantees exponential decay. Thus, asymptotic solutions with high mesonic quantum numbers $\ell' \gg \ell_{1,2}$ have tiny amplitudes,  and their contribution is effectively redundant. 
For more details, see \ref{sec_4body}.

\begin{figure}[t]
    \centering
    \includegraphics[width=0.6\textwidth]{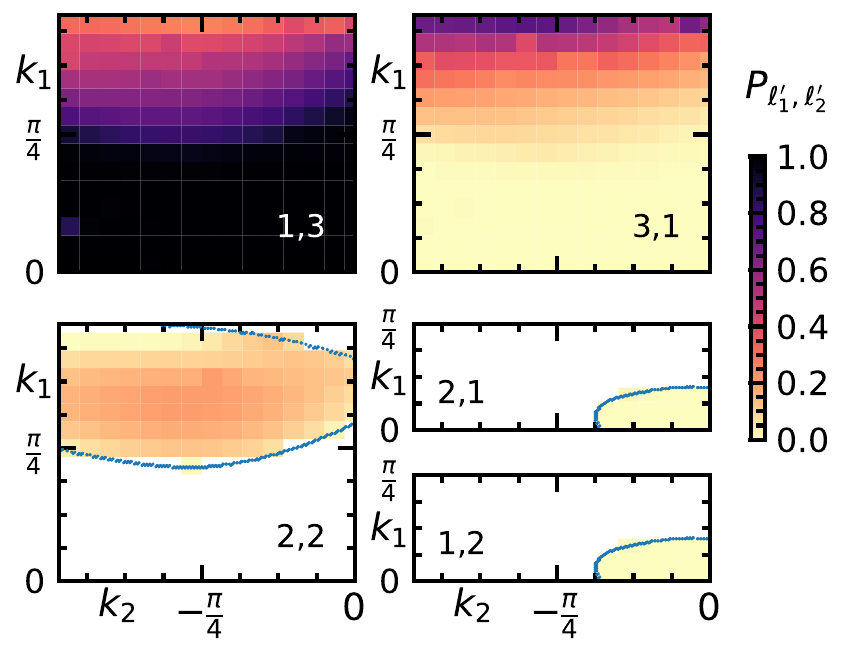}
    \caption{Probabilities of the various scattering channels $(1,3)\to(\ell'_1,\ell'_2)$ as a function of the incoming momenta, for $w/\tau=0.6$. The blue lines delimit the regions where the inelastic channels $(2,2)$, $(1,2)$, $(2,1)$ are open. {The probabilities of the channels plotted in the five panels sum up to one with good precision [small deviations from this value are shown in Fig.~\ref{fig:current}-(b)].}
    }
    \label{fig:probabilities}
\end{figure}

\mbox{In Fig.~\ref{fig:probabilities}} we plot the cross sections {$P_{\ell_1',\ell_2'}$} computed as described above, 
as a function of the incoming momenta $k_1$, $k_2$,
for the scattering  $(1,3)\to(\ell'_1,\ell'_2)$ 
when $w/\tau=0.6$. 

The scattering amplitudes can be readily connected with the products of real-time  wavepacket collisions. We verify this by numerically computing instances of the exact time evolution within the four-body subspace.
We consider the example in Fig.~\ref{fig:cartoon}:
{In a system} with $L=36$ fermionic sites, we prepare two Gaussian wavepackets $\Psi(s_1,s_2,r_1,r_2;t=0)=e^{-[(s_1-s_1^0)^2+(s_2-s_2^0)^2]/8\sigma^2}\psi_{\ell_1,k_1^0}(s_1,r_1)\psi_{\ell_2,k_2^0}(s_2,r_2)$ of the meson wavefunctions in Eq.~(\ref{eq_mesonwavefunction}) with $\ell_1=1$, $\ell_2=3$, centered around momenta $k_1^0=1.3$, $k_2^0=-0.4$ and positions $s_1^0=24$, $s_2^0=48$, with envelopes of width $\sigma=3\sqrt{2}$ lattice sites. 
Time evolution from this initial state is generated by the four-body Hamiltonian $H_{\text{4-body}}$ with $\tau=1$, $w=0.6$. 
The final state  at time $t_{f}=50$ is examined, when the wavepackets have collided and the products of the collision have not yet reached the  boundary of the system.
The energy and momentum of the incoming and outgoing states are represented in Fig.~\ref{fig:momentumdistribution}-(a). 
In Fig.~\ref{fig:momentumdistribution}-(b) and (c) we plot the joint
probability
distribution of the momenta $k_1$, $k_2$ at times $t=0$ and $t=t_{f}$, respectively, obtained via the Fourier transform of $\Psi(s_1,s_2,r_1,r_2;t)$ with respect to the center-of-mass positions $s_{1,2}$. While the initial state shows a single density peak at $(k_1^0, k_2^0)$, the final state gives three different density peaks, all lying on the line $k_1+k_2=k_1^0+k_2^0 \; {\rm mod}\; \pi$, demonstrating the conservation of total momentum. The three peaks 
correspond to the channels predicted from the {kinematic analysis} 
[the crosses in Fig.~\ref{fig:momentumdistribution}-(a)]: one for the trasmitted solution $(\ell_1',\ell_2')=(3,1)$ (with $k_1=-0.4$, $k_2=1.3$), one for the reflected solution $(1,3)$ ($k_1\simeq -1.0$, $k_2\simeq -1.2$), and one for the inelastic solution $(2,2)$ ($k_1\simeq -1.5$, $k_2\simeq -0.7$). The peaks can be better resolved by plotting the distribution of the relative momentum $k_1-k_2$, as done in Fig.~\ref{fig:momentumdistribution}-(e). 
The relative weights enclosed within the dashed contours in Fig.~\ref{fig:momentumdistribution}-(c),
$P^{\text{num}}_{13}\simeq 0.49$, $P^{\text{num}}_{22} \simeq 0.1$, $P^{\text{num}}_{31} \simeq 0.41$,
are compatible with the predicted cross sections \mbox{$P_{13}\simeq 0.47$, $P_{22}\simeq 0.12$, $P_{31}\simeq 0.41$}%
~\footnote{A smaller peak can be observed far from the momentum-conserving line $k_1+k_2=k_1^0+k_2^0$, and corresponds to the reflection of the second meson on the boundary after scattering in the $(3,1)$ channel. 
The missing probability fraction outside the dashed contours in Fig.~\ref{fig:momentumdistribution}-(c) is due to such effects as well as to the arbitrary cutoff used the define the contours, and amounts to $ \approx 20\%$ here.}.

\section{Quantum simulation}
\label{sec_quantumsim}
{The analysis above outlines a tractable regime where non-trivial meson scattering phenomena can be accessed and understood. We now discuss how to observe them --- and possibly extend their scope --- with a quantum simulator, which  minimally requires:} {\it i)} 
designing the desired Hamiltonian dynamics; {\it ii)} preparing the incoming state; {\it iii)} detecting the outgoing states.

\begin{figure}[t]
    \centering
    \includegraphics[width=0.6\textwidth]{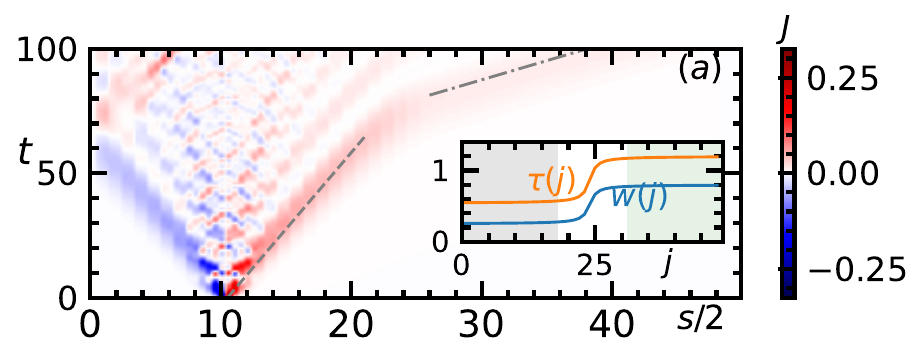}  
    \includegraphics[width=0.6\textwidth]{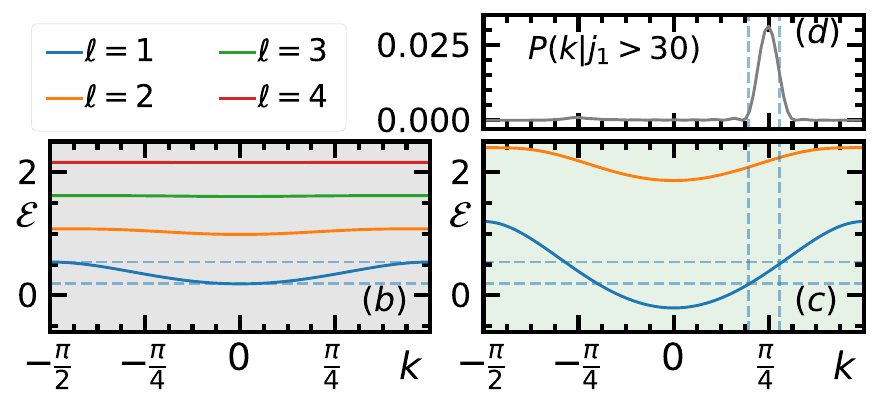}
    \caption{
    Manipulation of mesonic wavepackets by inhomogeneous fields. 
    (a) Time evolution of the meson current density $J(s, t)$ originating from a single spin flip (see the main text) at $s/2=(j_1+j_2)/2=10.5$, in the inhomogeneous field profiles shown in the inset, interpolating between $\tau_L=0.54$, $w_L=0.25$ (left) and $\tau_R=1.2$, $w_R=0.8$ (right). The slopes of the dashed and dashed-dotted lines correspond to the maximal velocity of the $\ell=1$ meson in the left and right regions, respectively. (b-c) Mesonic spectra in the left (b) and right (c) regions. The horizontal dashed lines indicate the range of allowed energies; the vertical lines define the range of momenta $k^*\pm \delta k$ allowed in the right region. (d) Momentum distribution of the transmitted meson wavepacket. 
    }
    \label{fig:preparation}
\end{figure}

Crucially, problem {\it i)} does not involve any experimental fine-tuning: the basic phenomena only rely on confinement, and are thus robust to any weak perturbation to the model.
As a concrete example,
we will focus on simulating
the Hamiltonian in Eq.~(\ref{eq_HZ2}) by exploiting the equivalence with the
quantum Ising chain in a tilted magnetic field~\cite{LeroseSuraceQuasilocalization,McCoyIsingLGT,Drouffe}:
\begin{equation}
\label{eq_Ising}
  { H_{\text{qIc}} = \sum_{j} -\frac m 2 \, \sigma^z_{j-1/2} \sigma^z_{j+1/2} + 
    \frac \tau 2  \, \sigma^z_{j+1/2} + 
    w \, \sigma^x_{j+1/2}.}
\end{equation}
To obtain this non-trivial identification, one has to exploit the gauge-symmetry constraints 
to eliminate the fermionic degrees of freedom~\cite{ZoharRemovingFermions,SuraceRydberg,LeroseSuraceQuasilocalization,borla2020gauging}, as reported in~\ref{sec_mapping}. 
Within this equivalence, $\mathbb{Z}_2$-mesons may be pictured as domains of reversed magnetization in a polarized background, arising from domain-wall confinement~\cite{McCoyWuConfinement,DelfinoMussardoSimonetti,DelfinoMussardo,ShankarConfinement,RutkevichMesonSpectrumLattice}. Recently, confinement in the quantum Ising chain has been shown to underlie a plethora of intriguing \mbox{non-equilibrium phenomena~\cite{Kormos:2017aa,MazzaTransport,LeroseSuraceQuasilocalization,RobinsonNonthermalStatesShort,Verdel19_ResonantSB}.}
The mapping above is extremely advantageous for quantum simulations because it implements gauge invariance exactly, {similarly to what done in Refs.~\cite{ExpPaper,SuraceRydberg} for the Schwinger model.}
 The dynamics governed by Eq.~(\ref{eq_Ising}) can be experimentally realized both with optical lattices~\cite{Greiner2011_TiltedMott,MeinertPRL2013_QuenchAtomicIsingChain} and Rydberg atoms trapped in optical tweezers~\cite{LeseleucRydberg,BernienRydberg}.

The preparation of the initial state \textit{ii)} {is subtle, as sharp meson wavepackets involve considerable entanglement between atoms, which is precluded to  single-site optical manipulations}.
We present here an approach exploiting spatially inhomogeneous fields 
in Eq.~(\ref{eq_Ising}) to filter meson wavepackets with sharply-defined momenta, {at the price of moderately longer chains and a limited amount of post-selection}. 
The numerical simulation in {Fig.~\ref{fig:preparation}-(a)} illustrates the core idea:
when $w/\tau \lesssim 1$, a spatially localized spin flip in the left region mostly excites the lowest (and fastest) meson $\ell=1$ at all momenta;
hence, a sharp spatial variation in the fields {$\tau(j),w(j)$} (inset) determines a corresponding change in the shape of mesonic bands [from that in panel (b) to that in (c)]; energy conservation (horizontal dashed lines) selects a narrow momentum window $k^*\pm \delta k$ (vertical dashed lines) for which rightward propagation is allowed. 
Panel~(d) shows that at time $t=50$ the fraction of mesonic wavepacket filtered in the right region is $\approx 20\%$ (the rest is reflected at the interface), and its momentum distribution has support within the selected window.
An analogous preparation can be made on the opposite side of the chain for the desired incoming mesonic wavepacket from the right.
Similarly, inhomogeneous fields can be used to accelerate mesons.

Finally, detecting the scattering products \textit{iii)} is conceptually simple, as the mesons involved in the various possible outcomes of a collision have different velocities [cf. Fig. \ref{fig:cartoon}], so they can be resolved as spatially separate wavepackets. 
For implementations {based on} Eq.~(\ref{eq_Ising}), 
the particle density $c^\dagger_j c_j$ in Eq.~(\ref{eq_HZ2}) maps to the domain-wall density {$(1-\sigma^z_{j-1/2} \sigma^z_{j+1/2})/2$}: Thus,
it is sufficient to measure the  magnetization profile $\langle \sigma^z_{j+1/2}(t_f) \rangle$ in the final state \cite{Greiner2011_TiltedMott,MeinertPRL2013_QuenchAtomicIsingChain,LeseleucRydberg,BernienRydberg} to reconstruct {the momenta of the mesons from their positions, the quantum numbers from their extension, and the cross sections from their probabilities.} 
We note that the required  time and length scales estimated from the above discussion ($50 \div 100$ lattice sites and units of time)  are within reach of present-day experiments: Ref.~\cite{BernienRydberg}, for example, demonstrated state preparation and single-qubit readout in a chain of $51$ {$^{87}$Rb} atoms governed by Ising-type dynamics close to Eq.~(\ref{eq_Ising}), with excellent coherence control over several tens of time units ($2\pi/w$).

\section{Outlook}
The analysis of the meson scattering problem and the proposed strategies for quantum simulations presented here can be straightforwardly applied to any one-dimensional model exhibiting confinement, including Abelian and non-Abelian lattice gauge theories (e.g., quantum link models~{\cite{SuraceRydberg,Chandrasekharan1997}}). 
They can also be extended to long-range interacting models, for which confinement effects \cite{GorshkovConfinement,LeroseDWLR} have been recently experimentally {investigated with trapped ions \cite{TanConfinementExperiment}.}
{The novel theoretical approach and exact solution to the meson scattering problem presented here will provide the basic building block for understanding the non-equilibrium evolution in quantum spin chains with confinement of excitations \cite{Kormos:2017aa,GorshkovConfinement}, particularly the recently reported lack of thermalization \cite{RobinsonNonthermalStatesShort,MazzaTransport,LeroseSuraceQuasilocalization,PaiPretkoFractonsLGT,Tagliacozzo2019,LeroseDWLR,BanulsWeakThermalization,LinMotrunichOscillations}.}

{Compared to real-world scattering experiments, quantum simulations naturally give access to full real-time resolution of the dynamics of a complex collision event, and to the pattern of quantum correlations and entanglement at the level of partons \cite{EPR_Ent,Kharzeev2017}, for which simplified lower-dimensional models such as the one discussed here could already provide deep insights.
In future work, we plan to investigate this, as well as to optimize schemes for cold-atom platforms.}
Intriguingly, quantum simulators could allow to explore regimes beyond our theoretical analysis such as the continuum limit \mbox{of quantum field theories \cite{FonsecaZamMesonSpectrumContinuum,RutkevichMesonSpectrumContinuum,SchwingerScatteringReview,Durr:2008ty,Rigobello}}. This would represent a first step towards the ultimate goal of simulating realistic scattering problems in QCD such as heavy-ion collisions \cite{WilczekCoherentOscillationsQCD}.

\ack
We acknowledge useful discussions and feedback by D. A. Abanin, A. Bastianello, P. Calabrese, M. Dalmonte,  W. De Roeck, G. Giudici, G. Pagano, J. Sonner, T. Wang, and our co-authors of Ref. \cite{LeroseSuraceQuasilocalization}. F.M.S. is partly supported by the ERC under grant number 758329 (AGEnTh).
A.L. is supported by the Swiss National Science Foundation.

\paragraph{Note added ---}
While completing the present manuscript, we became aware of a related work~\cite{Heyl_prep}, appeared simultaneously.

\appendix

\section{Gauge invariance and confinement}
\label{sec_gauge}

In this work we have focused on the $(1+1)$-dimensional $\mathbb{Z}_2$-LGT defined by the Hamiltonian in Eq. (1) of the main text, reported here for convenience:
\begin{equation}
\label{eq_HZ2sm}
    H= m \sum_j c^\dagger_{j} c_j \, +\frac{\tau}{2}  \sum_j  \sigma^z_{j+1/2} \\
    +w \sum_j 
\big(c_{j}^\dagger-c_j\big) \, \sigma^x_{j+1/2} \,
\big(c_{j+1} + c_{j+1}^{\dagger} \big).
\end{equation}
Equation (\ref{eq_HZ2sm}) describes  gauge-invariant interactions of fermionic particles mediated by a $\mathbb{Z}_2$ gauge field. The parameter $m$ represents the mass of the fermions, $\tau$ quantifies the excitation energy cost per site of the gauge field excitation (\textit{string tension}), and $w$ is the coupling strength.

Gauge-invariance entails that a fermion hop {or pair creation} across a bond $(j,j+1)$ is always accompanied by a flip of the gauge field $\sigma^z_{j+1/2}$ on that bond, as illustrated in Fig. \ref{fig_sketch}-(a). More formally, interactions are such that all the local operators
$
G_j=\sigma_{j-1/2}^z \, \sigma_{j+1/2}^z (1-2c^\dagger_{j}c_j)
$
are conserved, i.e., $[G_j,H]=0$. These operators satisfy $G_j^2 \equiv 1$ and thus generate local $\mathbb{Z}_2$ symmetries. 
Accordingly, the complete Hilbert space decomposes into dynamically disconnected subspaces, labelled by the set of eigenvalues $\{e^{i\pi q_{j}}=\pm 1\}$ of $\{G_{j}\}$, where $q_j=0$ or $1$ is interpreted as the absence or presence of a static background $\mathbb{Z}_2$-charge on site $j$, respectively.
In this work we have focused on the neutral gauge sector with $q_{j}\equiv0$, i.e., the subspace characterized by
\begin{equation}
\label{eq_gaugeneutral}
    \sigma_{j-1/2}^z \, \sigma_{j+1/2}^z = 1-2c^\dagger_{j}c_j
\end{equation}
for all $j$.
The meaning of this equation, referred to as the \textit{Gauss law}, is that the gauge field spatial variations can only take place upon crossing a site on which a fermion is located,
{as shown in Fig. \ref{fig_sketch}-(a).}

The LGT in Eq.~(\ref{eq_HZ2sm}) exhibits particle confinement in the regime {$m > 2|w|,|\tau|>0$}. 
The occurrence of fermion confinement in model~(\ref{eq_HZ2sm}) can be understood by considering the limit $m\to\infty$. The ground state $\ket{GS}$ becomes a fermion vacuum with $c^\dagger_j c_j \ket{GS} =0 $; by gauge-neutrality, [cf. Eq.~(\ref{eq_gaugeneutral})], 
the gauge field is uniformly polarized. 
The interaction term $H_{\text{int}}$ acting on the vacuum creates a pair of neighboring fermions. These particles can hop away from each other, with a hopping amplitude $w$.
As they move, gauge-invariance forces a string of excited gauge field {with tension $\tau$} to extend between them, which costs an energy per unit length equal to the string tension $\tau$. Therefore,the two $\mathbb{Z}_2$-charges experience a confining potential $V(r)=|\tau| r$ that grows unbounded at large distances $r$ whenever $\tau \neq 0$. As a consequence, the {excitations} form a discrete tower of \textit{neutral}  (in a $\mathbb{Z}_2$ theory, particle and antiparticle are the same object) bound states, labelled by their internal quantum number $\ell=1,2,\dots$ and their center-of-mass momentum $k$. 

\section{Mapping to the quantum Ising chain}
\label{sec_mapping}

\begin{figure}[t!]
    \centering
    \includegraphics[width=0.7\textwidth]{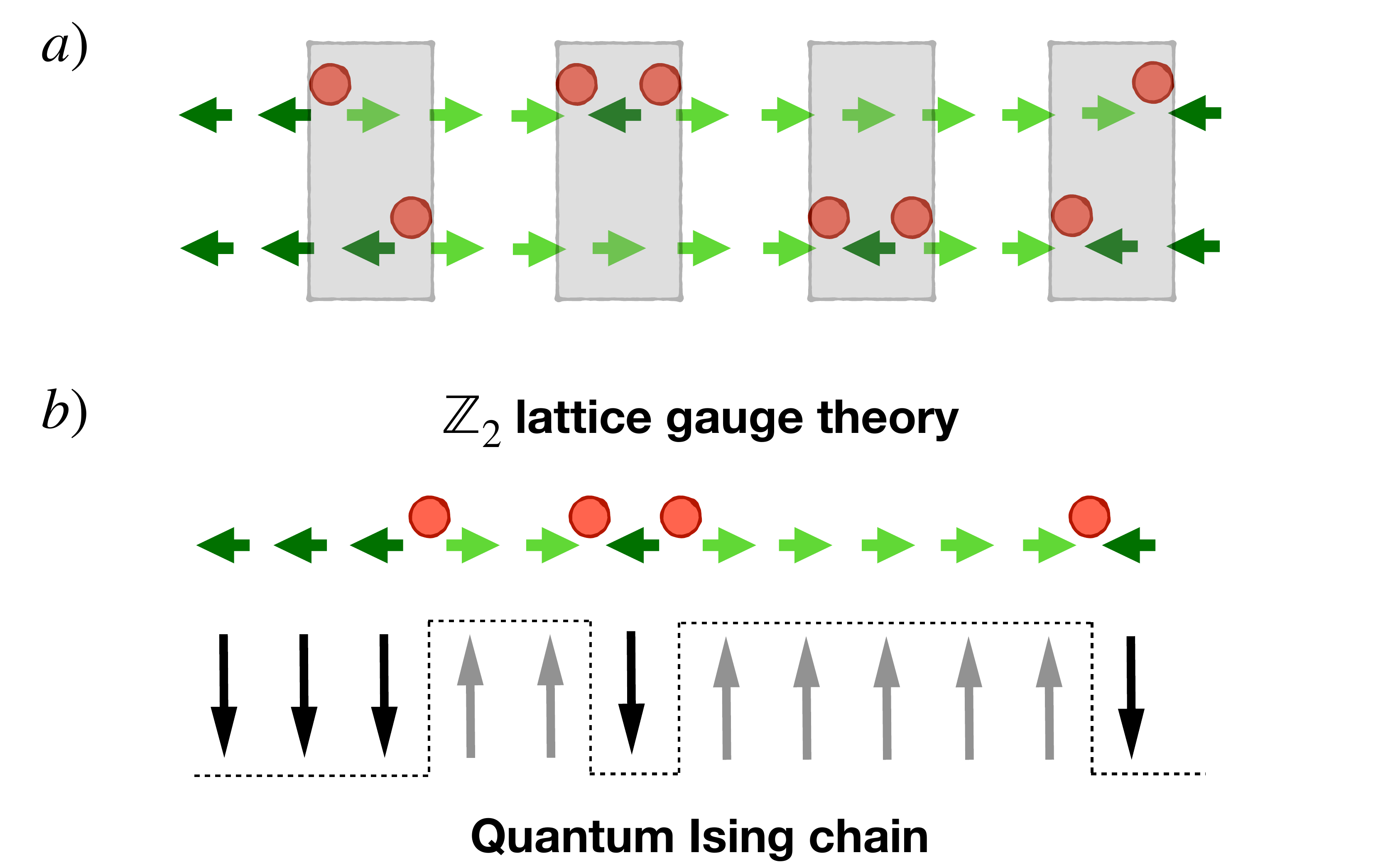}
    \caption{
    (a)  The two rows represent two gauge-neutral classical configurations of the matter and gauge fields in Eq. (\ref{eq_HZ2sm}), i.e., simultaneous eigenstates of the operators $n_j=c^\dagger_jc_j$ and $\sigma^z_{j+1/2}$ for all $j$.
    Red spots denote the presence of a fermionic charge;
    horizontal green arrows denote the polarization of the gauge field: light, rightward for $\sigma^z_{j+1/2}=+1$ and dark, leftward for $\sigma^z_{j+1/2}=-1$.
    By the Gauss law in Eq. (\ref{eq_gauss}), the gauge field varies only across sites that contain a charge.
    The grey-shaded rectangles highlight the possible local transitions, from the top to the bottom configuration, described by the interaction term in Eq. (\ref{eq_HZ2sm}): from left to right, we have rightward fermion hop, pair annihilation, pair creation, leftward fermion hop.
    (b) Cartoon illustration of the exact mapping between the $\mathbb{Z}_2$-LGT in Eq. (\ref{eq_HZ2sm}) in the neutral gauge sector defined by Eq.~(\ref{eq_gauss}), and the quantum Ising chain in Eq.~(\ref{eq_Ising}).
    The mapping hinges upon the elimination of the fermionic degrees of freedom, which are in one-to-one correspondence with gauge-field domain-walls by means of the Gauss law.
    }
    \label{fig_sketch}
\end{figure}

Here we illustrate the mapping between the $\mathbb{Z}_2$-LGT in Eq.~(\ref{eq_HZ2sm}) and the quantum Ising chain in a tilted magnetic field, Eq. (\ref{eq_Ising}).
This mapping was proposed in Ref.~\cite{LeroseSuraceQuasilocalization}, and is connected with the one discussed in Ref.~\cite{borla2020gauging}.

We consider the neutral sector, defined by the Gauss law (\ref{eq_gaugeneutral}), i.e., $G_j= \sigma_{j-1/2}^z \, \sigma_{j+1/2}^z (1-2n_j)\equiv 1$, 
with $n_j= c_j^{\dagger}c_j$ representing the number of fermions at site $j$.
The essence of the mapping is the exact elimination of the fermion degrees of freedom~\cite{ZoharCiracMatterIntegration}, as highlighted in Fig. \ref{fig_sketch}-(b). 
The latter are redundant due to the infinitely many local constraints in Eq.~(\ref{eq_gaugeneutral}):
a classical configuration of the gauge field uniquely fixes the configuration of the fermionic matter via the Gauss law.

We present a formal proof of the equivalence.
As a first step, we apply the Jordan-Wigner transformation to turn the fermions into hard-core bosons. To this aim, we introduce the spin-$1/2$ operators $\tau_j^\alpha$, for $\alpha=+,-,z$, defined as
\begin{equation}
    \tau_j^- = \prod_{k<j}(1-2n_k) c_j\, ,\qquad
    \tau_j^+=(\tau_j^-)^\dagger
    ,\qquad
    \tau_j^z = 2n_j-1 \, .
\end{equation} 
As can be easily checked, the operators $\tau_j^z$, $\tau_j^x= \tau_j^+ + \tau_j^- $ and $\tau_j^y= -i(\tau_j^+-\tau_j^-)$ satisfy the usual commutation relations of spin operators.
%
%
By applying this transformation, we get the Hamiltonian
\begin{equation}
    H= m \sum_j \frac{1+\tau_j^z}{2} \, +\frac{\tau}{2}  \sum_j  \sigma^z_{j+1/2} 
    +w \sum_j 
 \tau_j^x \, \sigma^x_{j+1/2} \,
\tau_{j+1}^x \, ,
\end{equation}
with the constraint \begin{equation}
\label{eq_gauss}
    G_j=-\sigma_{j-1/2}^z \, \sigma_{j+1/2}^z \tau_j^z=1.
\end{equation}

We now define a unitary transformation $U$ that eliminates the matter degrees of freedom. 
In other words, we seek for a
(gauge-variant) unitary $U$ such that the transformed Gauss law $G_j'=UG_jU^\dagger\equiv1$ from Eq.~(\ref{eq_gauss}) only depends on the matter degrees of freedom, whereas the transformed Hamiltonian $H'=UHU^\dagger$ only involves the gauge degrees of freedom. 
This can be accomplished with
\begin{eqnarray}
\fl
    U=\prod_j \exp\left[\frac{i\pi}{2}(\tau_{j}^x-1)\frac{1-\sigma_{j-1/2}^z\sigma_{j+1/2}^z}{2}\right]\nonumber\\
    =\prod_j \left[\frac{1+\sigma^z_{j-1/2}\sigma^z_{j+1/2}}{2}+\tau_j^x \frac{1-\sigma^z_{j-1/2}\sigma^z_{j+1/2}}{2}\right]. 
\end{eqnarray}
This transformation flips the spin $\tau_j^z$ where the neighbouring gauge fields are anti-aligned and does nothing where they are aligned. We get $U\tau_j^z U^\dagger =\tau_j^z\sigma_{j+1/2}^z\sigma_{j+1/2}^z$
and 
$U\tau_j^x \sigma^x_{j+1/2} \tau_{j+1}^x U^\dagger = \sigma^x_{j+1/2}$. The transformed constraint
\begin{equation}
\label{eq_transfGauss}
    G_j'=UG_jU^\dagger = -\tau_j^z\equiv 1
\end{equation}
forces the $\tau_j^\alpha$  spins to be polarized in the $-\hat z$ directions. 
They enter the transformed Hamiltonian only via $G_j'$: 
\begin{equation}
\label{eq_transfH}
    H'= m \sum_j \frac{1-G_j'\sigma_{j-1/2}^z  \sigma_{j+1/2}^z}{2} \, +\frac{\tau}{2}  \sum_j  \sigma^z_{j+1/2} 
    +w \sum_j 
 \sigma^x_{j+1/2} \,.
\end{equation}
Thus, in the neutral gauge sector, by Eq. (\ref{eq_transfGauss}) the spins $\tau_j^\alpha$ are eliminated. 
 Equation (\ref{eq_transfH}), which governs the dynamics within this sector, coincides with the quantum Ising chain in a tilted magnetic field reported in Eq.~(\ref{eq_Ising}) (up to an irrelevant additive constant).

We note that, while in this derivation we used a non-local transformation to convert the fermions into hard-core boson, it is nevertheless possible to formulate a completely local mapping between the two Hamiltonians: in the neutral gauge sector, the Jordan-Wigner string can be completely reabsorbed using Gauss' law. To show that the mapping is local, it is sufficient to define the transformed spin operators
\begin{eqnarray}
    \tilde \sigma_{j+1/2}^x=(c_j^\dagger-c_j)\sigma^x_{j+1/2}(c_{j+1}^\dagger+c_{j+1}),\\
    \tilde \sigma_{j+1/2}^y=(c_j^\dagger-c_j)\sigma^y_{j+1/2}(c_{j+1}^\dagger+c_{j+1}),\\
    \tilde \sigma_{j+1/2}^z=\sigma^x_{j+1/2}.
\end{eqnarray}
These operators satisfy the usual commutation relations of Pauli matrices and are related to the original spins by a local transformation. It is then immediate to write Eq.~(\ref{eq_HZ2sm}) in terms of the new spin operators, obtaining the quantum Ising chain in the neutral gauge sector:
\begin{equation}
    H'= m \sum_j \frac{1-\tilde\sigma_{j-1/2}^z  \tilde\sigma_{j+1/2}^z G_j}{2} \, +\frac{\tau}{2}  \sum_j  \tilde\sigma^z_{j+1/2} 
    +w \sum_j 
 \tilde\sigma^x_{j+1/2} \,.
\end{equation}

\section{Solution of the two-body problem}
\label{sec_2body}

The two-body Hamiltonian for $m\to\infty$ is obtained by projecting Eq. (\ref{eq_HZ2sm}) onto the two-fermion subspace. It can be written in the basis of the fermion positions as
\begin{eqnarray}
\label{eq_twobody}
\fl
    H_{\text{2-body}} = \sum_{j_1<j_2}  \tau (j_2-j_1) \;  \ket{j_1,j_2} \bra{j_1,j_2}
    \nonumber\\
    + w \Big(
    \ket{j_1+1,j_2} \bra{j_1,j_2}+
    \ket{j_1,j_2+1} \bra{j_1,j_2}+ \text{H.c.}
    \Big) \, ,
\end{eqnarray}
where $j_1,j_2$ label the positions of the two fermions along the chain.

For $\tau=0$ the eigenstates are (antisymmetric combinations of) plane waves 
\begin{equation}
    \Psi_{k_1,k_2}=e^{i k_1j_1+ik_2j_2}-e^{i k_2j_1+ik_1j_2}
\end{equation} 
with energy $\mathcal{E}_{\text{free}}(k_1)+\mathcal{E}_{\text{free}}(k_2)$, where $\mathcal{E}_{\text{free}}(k) = 2 w \cos k$, $k\in[-\pi,\pi)$ is the free-fermion dispersion relation. For $\tau\neq0$, however, a linear confining potential emerges between the two fermions, and the spectrum is nonperturbatively modified into a discrete tower of bound states (mesons) labelled by a quantum number $\ell=1,2,\dots$. Each meson has a different dispersion relation $\mathcal{E}_\ell(k)$, where $k$ is the center-of-mass momentum of the bound state.

The meson wavefunctions and dispersion relations can be solved explicitly \cite{FogedbyTwoKinkSolution,LeroseSuraceQuasilocalization} by switching to the center-of-mass and relative variables, $s=j_1+j_2$, $r=j_2-j_1>0$. Substituting the ansatz
\begin{equation}
    \psi_k(s,r) = e^{iks} \phi_k(r)
\end{equation}
into the Schroedinger equation, we get a Wannier-Stark equation for the reduced wavefunction $\phi_k(r)$,
\begin{equation}
\label{eq_reduced2body}
    \tilde{w}_k \big[\phi_k(r+1)+\phi_k(r-1)\big] + \tau r \, \phi_k(r) = \mathcal{E} \, \phi_k(r)
\end{equation}
with an effective hopping $\tilde{w}_k=2w\cos k$ and, crucially, the boundary condition $\phi_k(0)\equiv 0$ due to Pauli exclusion.
This equation is equivalent to the recursion relation of the Bessel functions:
\begin{equation}
    \phi_k(r) = \mathcal{J}_{r-\mathcal{E}/\tau}(2\tilde{w}_k/\tau).
\end{equation}
The boundary condition $\mathcal{J}_{-\mathcal{E}/\tau}(2\tilde{w}_k/\tau)=0$ yields the quantization rule
\begin{equation}
\label{eq_mesonspectrum}
    \mathcal{E} = \mathcal{E}_\ell(k)=\tau \, \nu_\ell(2\tilde{w}_k/\tau) \equiv - \tau \times \text{ \{$\ell$-th zero of $x \mapsto \mathcal{J}_x(2\tilde{w}_k/\tau)$\} }
\end{equation}
for $\ell=1,2,\dots$,
which defines the exact dispersion relations of all mesons.
The meson wavefunctions are thus
\begin{equation}
\label{eq_mesonwavefunction}
    \psi_{\ell, k}(s,r) = e^{iks} \mathcal{J}_{r-\nu_\ell}(2\tilde{w}_k/\tau).
\end{equation}
Note that $k\in[-\pi/2,\pi/2)$, because $k$ and $k+\pi$ generate the same solution up to a phase: Since $\mathcal{J}_{\alpha}(-z)=e^{i\pi \alpha}\mathcal{J}_{\alpha}(z)$, when $k\mapsto k+\pi$ the wavefunction $\psi$ gets multiplied by $(-)^s e^{i\pi(r-\nu_\ell)}=e^{-i \pi \nu_\ell} (-)^{2 j_2}=e^{-i \pi \nu_\ell}$, i.e., a global phase.

The most important qualitative aspects of this exact solution are the following. For $w\to0$, one finds 
energies
$\mathcal{E}_\ell(k)=\tau \ell$, corresponding to a 
pair of fermions separated by a string of excited gauge fields
of length $\ell$. In this limit, bound states are dispersionless (flat bands). As $w$ increases, the lightest mesons progressively acquire mobility (band curvature). In particular, one can see that an effective hopping of the $\ell$-th meson appears at the $2\ell$-th order in perturbation theory in $w/\tau$, which gives rise to a band curvature (and hence a maximal velocity) of this order of magnitude. This can be confirmed by the exact solution above, as \cite{Abramowitz} 
\begin{equation}
\label{eq_Besselzeroexpansion}
    \nu_\ell(x) \; \underset{x\to0}{\thicksim} \; \ell \bigg(1 - \frac {x^{2\ell}}{(\ell!)^2} \bigg) \; .
\end{equation}
Interestingly, the flat-band property of heavy mesons is a nonperturbative feature that persists to arbitrarily large values of the ratio $w/\tau$. In fact, for $\ell\gg 4w/\tau$, the band curvature drops to zero faster than exponentially. 
This phenomenon is due to Wannier-Stark localization of particles in a linear potential: Single, isolated particles can be seen to perform a finite oscillatory motion (Bloch oscillations) with an amplitude of $\xi=2w/\tau$ lattice sites. Correspondingly, their eigenstates are localized around each lattice site, with a localization length $\xi$.
When two particles are initialized at a distance much larger than $2\xi$, they perform independent oscillations without touching each other, and the meson is thus immobile and localized. The mobility is provided by the hard-core interaction between the two kinks, which is suppressed as the overlap between the two localized wavefunction tails, corresponding to the estimate in Eq. (\ref{eq_Besselzeroexpansion}) \cite{LeroseSuraceQuasilocalization}.

There exist solutions of the Schroedinger equation (\ref{eq_reduced2body}) with complex momentum $k$ and energy $\mathcal{E}$, with the same wavefunction (\ref{eq_mesonwavefunction}) and the same (analytically continued) energy-momentum relation (\ref{eq_mesonspectrum}). Such solutions correspond to evanescent waves and are important in the scattering problem that will be analyzed below.

\section{Solution of the four-body problem}
\label{sec_4body}

The four-body Hamiltonian for $m\to\infty$ is obtained by projecting Eq. (\ref{eq_HZ2sm}) onto the four-fermion subspace. It can be written in the basis of the fermion positions as
\begin{equation}
\label{eq_fourbody}
\fl
    H_{\text{4-body}} = \sum_{j_1<j_2<j_3<j_4}
    \Bigg[\tau(j_2-j_1+j_4-j_3) \; \ket{\vec j} \bra{\vec j}
      + w \sum_{n=1}^4 \Big(
    \ket{\vec j +\hat e_n} \bra{\vec j}+\text{H.c.}
    \Big)\Bigg]
\end{equation}
where $\vec{j}=(j_1,j_2,j_3,j_4)$ and we defined the unit vectors $\hat e_1=(1,0,0,0)$, $\hat e_2=(0,1,0,0)$, etc.
The 
sum is constrained by Pauli exclusion.

The diagonal term $\propto \tau$ can be viewed as the two confining potentials for the first and last pair of adjacent fermions.
These potentials give rise to two bound states (mesons). These mesons experience no residual interactions; they only interact when they bump into each other, due to Pauli exclusion.
For later convenience, we define the center-of-mass positions $s_1=j_1+j_2$, $s_2=j_3+j_4$ and relative distances $r_1=j_2-j_1$, $r_2=j_4-j_3$ for the two mesons.

We are interested in the problem of a scattering event with incoming mesons in states $(\ell_1,k_1)$, $(\ell_2,k_2)$. This asymptotic state defines the total energy $E\equiv \mathcal E_{\ell_1}(k_1)+\mathcal E_{\ell_2}(k_2)$
and the total momentum 
$K \equiv k_1+k_2\; {\rm mod}\;\pi$ of the system.
Since meson-meson interactions are local, we formulate an ansatz in terms of the product state
\begin{equation}
   \chi_{\ell_1', k_1', \ell_2', k_2'}(s_1,r_1,s_2,r_2)= \psi_{\ell_1', k_1'}(s_1,r_1) \psi_{\ell_2', k_2'}(s_2,r_2),
\end{equation}
where $\psi_{\ell_{1,2}', k_{1,2}'}$ are eigenstates of the two-body problem with quantum numbers $\ell_{1,2}'$ and generally complex momenta $k_{1,2}'\in \mathbb{C}$. 
For $\Im (k_1')\le 0$, $\Im (k_2')\ge 0$, the ansatz $\chi_{\ell_1', k_1', \ell_2', k_2'}$ is an asymptotic solution, and solves the Schr\"{o}dinger equation in the full domain $(s_2-s_1-r_1-r_2)/2=j_3-j_2>0$ away from the scattering region (that is the hyperplane $j_2=j_3$).

To obtain the complete solution for given scattering data, we first need to determine the set of parameters $\{(\ell_1^\alpha, k_1^\alpha), (\ell_2^\alpha, k_2^\alpha)\}_{\alpha =1,2,\dots}$ which simultaneously satisfy the conservation laws of total energy and momentum
\begin{equation}
    E= \mathcal E_{\ell_1^\alpha}(k_1^\alpha)+\mathcal E_{\ell_2^\alpha}(k_2^\alpha),
\end{equation}
\begin{equation}
    K =k_1^\alpha+k_2^\alpha\; {\rm mod}\; \pi.
\end{equation}
Note that, while $E$ and $K$ are real, $k_{1,2}^\alpha$ and $\mathcal E_{\ell_{1,2}^\alpha}(k_{1,2}^\alpha)$ are generally complex. 
Solutions with real momentum and energy correspond to incoming or outgoing states, are only a finite number.
The candidate scattering solution is a linear superposition of the form
\begin{equation}
\label{eq:superpos}
    \Psi(s_1,r_1,s_2,r_2)=\chi_{\ell_1,k_1,\ell_2,k_2}+\sum_\alpha A_\alpha \chi_{\ell_1^\alpha,k_1^\alpha,\ell_2^\alpha,k_2^\alpha}.
\end{equation}
{The sum in Eq.~(\ref{eq:superpos}) has to be restricted to the asymptotic solutions with outgoing current (see below) among those with real energy/momentum, and includes all the evanescent states that decay exponentially with the distance from the scattering region 
among those with complex energy/momentum.}

The values of the coefficients $A_\alpha$ are then obtained by imposing the continuity of the solution in the scattering region. More explicitly, we have to impose that $\Psi \equiv 0$ on the hyperplane $j_2=j_3$. We define
$ S = (s_1+s_2)/2$, $R =s_2-s_1$, representing the global center-of-mass position and the distance between the two mesons, respectively. Hence, we have
\begin{eqnarray}
\label{eq:sol}
\fl
    \Psi(S, R,r_1,r_2)=e^{iKS}\Big[e^{ipR}\phi_{\ell_1,K/2-p}(r_1)\phi_{\ell_2,K/2+p}(r_2)\nonumber\\
    +\sum_\alpha A_\alpha e^{ip_\alpha R}\phi_{\ell_1^\alpha,K/2-p_\alpha}(r_1)\phi_{\ell_2^\alpha,K/2+p_\alpha}(r_2)\Big]
\end{eqnarray}
where $p=(k_2-k_1)/2$, $p_\alpha=(k_2^\alpha-k_1^\alpha)/2$, and $\phi_{\ell, k}$ are the solutions to Eq.~(\ref{eq_reduced2body}). The boundary condition applied to Eq.~(\ref{eq:sol}) reads
\begin{equation}
\label{eq:linsys}
    \sum_\alpha M_{(r_1,r_2), \alpha} A_\alpha =B_{(r_1,r_2)},
\end{equation}
where
\begin{equation}
    M_{(r_1,r_2), \alpha}= e^{ip_\alpha (r_1+r_2)}\phi_{\ell_1^\alpha,K/2-p_\alpha}(r_1)\phi_{\ell_2^\alpha,K/2+p_\alpha}(r_2),
\end{equation}
\begin{equation}
    B_{(r_1,r_2)}=- e^{ip(r_1+r_2)}\phi_{\ell_1,K/2-p}(r_1)\phi_{\ell_2,K/2+p}(r_2).
\end{equation}

The coefficients $A_\alpha$ are then obtained by solving the linear system in Eq.~(\ref{eq:linsys}), truncated to a finite set of values of $\alpha$ and $r_{1,2}\le r_{\text{max}}$. 
{The truncation in $\alpha$ can be safely performed:
the open outgoing channels are only a finite number, and the evanescent states with increasingly high quantum numbers have large $\Im (p_\alpha)$, resulting in negligible contributions. 
The truncation to $r_1,r_2\le r_{\text{max}}$ is also legitimate:
$B_{(r_1,r_2)}$ decays exponentially fast with $r_1$ and $r_2$, thanks to the spatial decay of the mesonic wavefunctions $\phi_{\ell,k}$; the coefficients
$M_{(r_1,r_2), \alpha}$ decay for the same reason when $\alpha$ represents an outgoing solution with $p\in\mathbb{R}$, whereas the normalizability condition $\Im(p)>0$ guarantees the decay of the prefactor $e^{ip_\alpha(r_1+r_2)}$ when $\alpha$ represents an evanescent solution.}
In all the calculation presented in the main text, we have checked convergence with respect to these truncation cutoffs.

\section{Mesonic current}
\label{sec_current}


{For sufficiently large $m$ (see \ref{sec_finitemass}),} in a scattering process, the number of mesons is globally conserved. This conservation law is associated with a continuity equation.
We now illustrate this continuity equation for the generic case of $q$ mesons (i.e., in the $2q$-fermion sector) in limit $m\rightarrow \infty$. We define the density operator for the $i$-th meson at position $x$
\begin{equation}
 \chi_i(x) = \sum_{j_1<j_2<\dots<j_{2q}}  \delta_{j_{2i-1}+j_{2i} \, , \, x} \ket{\vec j} \bra{\vec j}  
\end{equation}
with $\vec j=(j_1,j_2,\dots,j_{2q})$,
and the total mesonic density 
\begin{equation}
    \rho(x)=\sum_{i=1}^q \rho_i(x) \equiv \sum_{i=1}^q \bra{\Psi}\chi_i(x)\ket{\Psi}\, .
\end{equation} 
The mesonic current is defined as
$J (x) =\sum_{i=1}^{q} J_i(x)$, with
\begin{equation}
    \fl J_i(x) =-2w\sum_{j_1<j_2<\dots<j_{2q}} \delta_{j_{2i}+j_{2i+1}\, , \, x}
    \; \Im \left\{ \Psi^*(\vec j) \left[\Psi(\vec j+\hat e_{2i})+\Psi(\vec j +\hat e_{2i+1})\right]\right\},
\end{equation}
where $\hat e_n$ is the unit vector along the direction of $j_n$.
\begin{figure}[t!]
    \centering
    \includegraphics[width=0.4\textwidth]{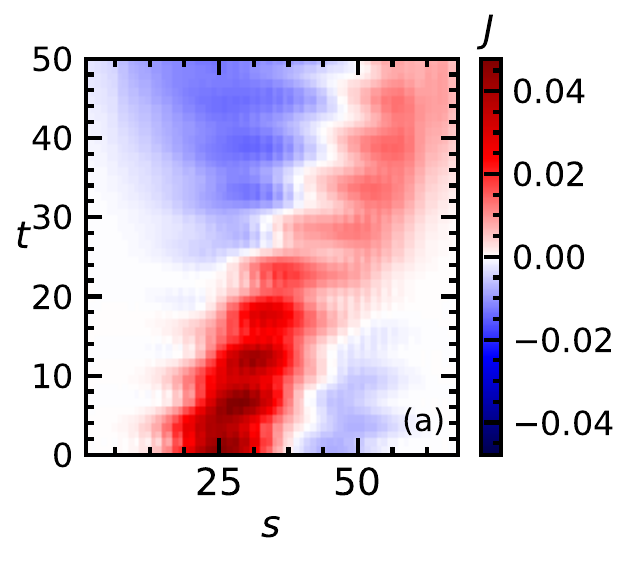}
    \includegraphics[width=0.36\textwidth]{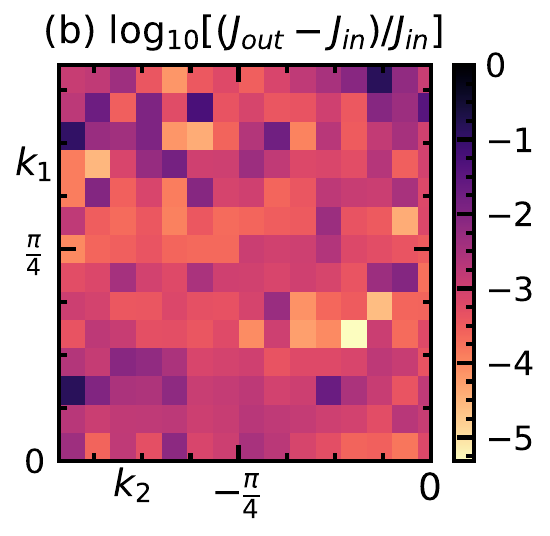}
    \caption{(a) Profile of $J(x,t)$ in a scattering event. The parameters of the simulation and the initial state are the same as in Fig. \ref{fig:momentumdistribution}. (b) Violation of current conservation law in the analytical solution of the scattering problem. This quantity serves as a consistency check for the theory. Except for a small number of points (for which we attribute the inaccuracy to numerical instabilities in determining the set of solutions), the observed error in the current conservation is  small.}
    \label{fig:current}
\end{figure}
As an example, in Fig.~\ref{fig:current}-(a) we plot the time-evolving profile of the meson current $J(x,t)$ for the scattering event discussed in Fig.~(\ref{fig:momentumdistribution}). The red (blue) color is associated to positive (negative) current, i.e., to a meson moving to the right (left). 

The mesonic density and current satisfy the continuity equation
\begin{equation}
\label{eq:conteq}
    \partial_t \rho(x) = -J(x)+J(x-1).
\end{equation}
The proof of this equation is reported in the next section.
We now derive the constraints imposed on the scattering solutions of the Schroedinger equation by the continuity equation.

Let us first consider the case of a single meson, with internal quantum number $\ell$ and momentum $k$. The density current associated with it can be written in terms of its dispersion relation $\mathcal{E}_\ell(k)$ and wavefunction $\phi_{\ell,k}(r)$ as
\begin{eqnarray}
    J(x)&= -2w \Im \left\{ \sum_{r>0} \phi_{\ell, k}^*(r) e^{ik}[\phi_{\ell, k}(r+1)+\phi_{\ell, k}(r-1)]\right\}\nonumber\\
    &= \sum_{r>0} \phi_{\ell, k}^*(r) (-2w \sin k)[\phi_{\ell, k}(r+1)+\phi_{\ell, k}(r-1)]\nonumber\\
    &=\bra{\phi_{\ell, k}}\partial_k H_k \ket{\phi_{\ell,k}}=\partial_k \mathcal{E}_\ell(k)=v_\ell(k).
\end{eqnarray}
{where
\begin{equation}
    H_k =\sum_{r>0} 2w\cos k \big(\ket{r}\bra{r+1}+\text{H.c.}\big)+\tau r \ket{r}\bra{r}
\end{equation}
is the reduced Hamiltonian for the internal coordinate of the meson with center-of-mass momentum $k$ [see Eq.~(\ref{eq_reduced2body})].}
We obtain that the mesonic current corresponds to the group velocity of the meson, in analogy with the case of a structureless quantum particle.

We now apply the continuity equation to the solution $\Psi(s_1,r_1,s_2,r_2)$ of the stationary Schr\"{o}dinger equation for the scattering problem discussed in the main text. Equation~(\ref{eq:conteq}) implies that $J_{R}=J_{L}$, where $J_R$ and $J_L$ are the currents on the right and on the left, very far from the scattering region. There, the density current can be easily computed as the sum of the currents of the isolated mesons. Since evanescent waves do not contribute far from the scattering region, only propagating waves (i.e., those with $q_\alpha \in \mathbb{R}$) should be taken into account. Therefore, in a scattering process with incoming mesons of quantum numbers $\ell_1$ and $\ell_2$ and momenta $k_1$ and $k_2$, the two currents read
\begin{equation}
    J_{L}=v(\ell_1, k_1)+\sum_{\alpha|q_\alpha \in \mathbb{R}} |A_{\alpha}|^2 v(\ell_1^\alpha, k_1^\alpha)
\end{equation}
\begin{equation}
    J_{R}=v(\ell_2, k_2)+\sum_{\alpha|q_\alpha \in \mathbb{R}} |A_{\alpha}|^2 v(\ell_2^\alpha, k_2^\alpha).
\end{equation}
where $k_1^\alpha=q_\alpha$, $k_2^\alpha=K-q_\alpha$ are the momenta of the outgoing mesons.

The condition $J_R=J_L$ can be equivalently formulated as an equality between the incoming and outgoing currents $J_{in}=J_{out}$, defined as
\begin{equation}
    J_{in}=v(\ell_1, k_1)-v(\ell_2, k_2)
\end{equation}
and
\begin{equation}
    J_{out}=\sum_\alpha J_\alpha=\sum_\alpha |A_{\alpha}|^2[-v(\ell_1^\alpha, k_1^\alpha)+v(\ell_2^\alpha, k_2^\alpha)].
\end{equation}
The equation $J_{in}=J_{out}$ has an immediate physical interpretation as a conservation of probability: in a scattering event, at $t=-\infty$ the two mesons are with probability 1 in the state $\{(\ell_1, k_1), (\ell_2, k_2)\}$; at $t=+\infty$, the outgoing meson states $\{(\ell_1^\alpha, k_1^\alpha), (\ell_2^\alpha, k_2^\alpha)\}$ have fractional probabilities $P_\alpha=J_\alpha/J_{in}$. 
Similarly to the scattering of structureless quantum particles, the probability of finding a certain scattering outcome (or total cross section) 
is proportional to the width of the wavepacket, which is determined by both the squared amplitude $|A_\alpha|^2$ and the group velocity. 

We stress that the sign of the total current defines outgoing states, characterized by $-v(\ell_1^\alpha, k_1^\alpha)+v(\ell_2^\alpha, k_2^\alpha)>0$. In computing the amplitudes of a scattering event, one has to select the set of propagating asymptotic solutions according to this criterion, as anticipated in \ref{sec_4body} above. 

We finally note that the conservation law $J_{in}=J_{out}$ represents a consistency check on our results for the coefficients $A_\alpha$ obtained from the truncation of the linear system in Eq. (\ref{eq:linsys}). In Fig.~\ref{fig:current}-(b) we plot the relative violation of this conservation law, for the computations involved in Fig. \ref{fig:probabilities}.
\subsection{Proof of the continuity equation}
\label{sec_continuity}

We prove here the continuity equation (\ref{eq:conteq}).

In the sector with $q$ mesons, we define the operators $\Delta^+=\sum_{i=1}^{q}\Delta_i^+$ where
  \begin{equation}
     \Delta_i^+=
    \sum_{j_1<j_2<\dots<j_{2q}}\sum_{s=2i,2i+1}  w(1-\delta_{j_{s+1},j_{s}+1}) \ket{\vec j+\hat e_s}\bra{\vec j}.
\end{equation}
The Hamiltonian can be written as
$H=\Delta^+ +\Delta^- + V$, where $\Delta^-=(\Delta^+)^\dagger$, $V=\sum_i V_i$  and
\begin{equation}
    V_i = \sum_{j_1<j_2<\dots<j_{2q}} \tau (j_{2i+1}-j_{2i})\ket{\vec j}\bra{\vec j}.
\end{equation}
The Heisenberg evolution of the meson density reads
\begin{eqnarray}
    \partial_t \rho (x) &=i\sum_{i=1}^q\bra{\Psi}[H, \chi_i(x)]\ket{\Psi}\nonumber\\
    &=i\sum_{i=1}^q\bra{\Psi}[\Delta_i^++\Delta_i^-, \chi_i(x)]\ket{\Psi}\nonumber\\
    &=2 \sum_{i=1}^q \Im \bra{\Psi}\chi_i(x)(\Delta^+_i + \Delta^-_i)\ket{\Psi}.
\end{eqnarray}
By using the properties $\Delta^-_i\chi_i(x)=\chi_i(x-1)\Delta^-_i$ and $\bra{\Psi}\Delta^-_i\chi_i(x) \ket{\Psi}=(\bra{\Psi}\chi_i(x) \Delta^+_i \ket{\Psi})^*$ we get
\begin{eqnarray}
\partial_t \rho(x) & = 2 \sum_{i=1}^m \Im \bra{\Psi}\chi_i(x)\Delta^-_i + \chi_i(x)\Delta^+_i \ket{\Psi}\nonumber\\
    & = 2\sum_{i=1}^m \Im \bra{\Psi}\chi_i(x)\Delta^-_i - \Delta^-_i\chi_i(x) \ket{\Psi}\nonumber\\
   & = 2\sum_{i=1}^m \Im \bra{\Psi}(\chi_i(x)-\chi_i(x-1))\Delta^-_i \ket{\Psi}\nonumber\\
    & =-J(x)+J(x-1) \, ,
\end{eqnarray}
i.e., Eq. (\ref{eq:conteq}).

\section{Finite fermion mass}
\label{sec_finitemass}

For the sake of simplicity, the discussion above and in the main text focuses on the limit $m\to\infty$.
We compactly summarize here the effects of a finite fermion mass.

\paragraph{Perturbative corrections to the exact spectra and scattering solution --- }
The main consequence of the finiteness of the fermion mass $m$ is to produce a perturbative dressing of the vacuum and of the excitations. 
These effects can be explicitly computed order by order via  the so-called Schrieffer-Wolff transformation~\cite{FrohlichSchriefferWolff,LossSchriefferWolff,McDonaldSchiefferWolffHubbard}. 
In this scheme, one sequentially solves for unitary transformations 
$U_n=e^{i m^{-n} S_{n}}\dots e^{i m^{-1} S_{1}}$, where the \mbox{$n$-th} generator $S_n$ is chosen to exactly cancel all processes violating fermion number conservation, in such a way that the transformed Hamiltonian $H'_n = U_n H U_n^\dagger$ at the $n$-th step commutes with $H_0=\sum_j c^\dagger_j c_j$ up to terms of order $m^{-n}$:
\begin{equation}
H'_n =  m H_0 + H_1 + m^{-1} H_2 + \dots + m^{-n+1} H_n +  \mathcal{O}(m^{-n})
,
\end{equation}
with $[H_n,H_0]=0$ for all $n$. The approximate Hamiltonian obtained by neglecting the higher-order remainder conserves the total fermion number, and exactly accounts for all perturbative $n$-th order transitions within each fermion-number sector  occurring through up to $n$ 
 {virtual} transitions involving intermediate states in other sectors.
 
Upon restricting the transformed Hamiltonian to the $2q$-fermion sector, one ends up with higher-order corrections to $H_{\text{2q-body}}$. 
For example, the first correction involves next-nearest-neighbor particle hopping terms with amplitudes $w^2/2m$:
\begin{eqnarray}
\label{eq_perturb}
\fl    H_{\text{$2q$-body}}^{(1)} = \sum_{j_1<\dots<j_{2q}}
    \Bigg[\tau\sum_{n=1}^{2q}(-)^n j_n \; \ket{\vec j} \bra{\vec j}
      + w \sum_{n=1}^{2q} \Big(
    \ket{\vec j +\hat e_n} \bra{\vec j}+\text{H.c.}
    \Big)\nonumber\\
    - \frac{w^2}{2m} \sum_{n=1}^{2q} \Big(
    \ket{\vec j +\hat 2 e_n} \bra{\vec j}+\text{H.c.}
    \Big)
    \Bigg]
\end{eqnarray}
with the understanding that 
\begin{eqnarray}
    \ket{j_1,\dots,j_n=j_{n+1},\dots,j_{2q}} \equiv 0, \\
    \ket{j_1,\dots,j_n=j_{n+1}+1,\dots,j_{2q}} \equiv 
    -\ket{j_1,\dots,j_{n+1},j_n,\dots,j_{2q}}.
\end{eqnarray}
Similarly, corrections of order $w^{r}/m^{r-1}$ introduce new hopping terms of range $r$ and renormalize shorter-range terms.

From the perturbatively corrected Hamiltonian, we can in principle derive the mesonic spectra, the scattering amplitudes and the mesonic currents to arbitrarily good accuracy, as long as $m\gg |w|$.

\paragraph{Particle pair creation in high-energy collisions --- }
In the regime considered in this work, fermionic pair creation is energetically forbidden, because the fermion mass $\sim m$ exceeds by far the kinetic bandwidth of excitations $\sim w$.
However, this phenomenon becomes relevant  when $m \simeq 2 |w|$.
This can be inferred from the exact spectrum of the free fermions for $\tau=0$, obtained from the equivalence with the solvable transverse-field Ising chain (see \ref{sec_mapping}):
\begin{equation}
\mathcal{E}_{\text{free}}(k)= m \sqrt{1+\frac{4w^2}{m^2} + 4 \frac w m \cos k }.
\end{equation}
When $m$ approaches $2|w|$ (from above), the renormalized mass $\mu \equiv \min_k \mathcal{E}(k)=m-2|w|$ of fermionic particles decreases to small values, and the bandwidth is $\sim 2 m$. 
Thus, if a weak string tension $\tau\neq0$ is considered, the kinetic energy of mesons 
can reach values much larger than their rest mass $\sim 2\mu$, and thus high-energy collisions could generate extra mesons.
This phenomenon goes beyond the theoretical analysis presented in this work, but could be accessed with quantum simulators.

\paragraph{Decay of heavy mesons --- }
A finite fermion mass may also trigger the instability of heavy mesons, which can decay into two or more lighter mesons when their gauge field string is sufficiently extended (\textit{string breaking}). 
The lifetime of unstable mesons is (at least) exponentially long in the ratio $m/|w|$, as discussed in Ref. \cite{LeroseSuraceQuasilocalization}; James et al. \cite{RobinsonNonthermalStatesShort} have argued that it may even be infinite, based on numerical evidence.
Thus, this phenomenon is not relevant in the regime studied in this work.
The instability threshold is instead relevant when approaching the continuum limit $m \searrow 2|w|$, where the model exhibits an emergent Lorentz invariance (as can be inferred from the exact mapping in \ref{sec_mapping}). In this regime, the lifetimes of mesons with mass $M>4\mu$ is only perturbative  $\sim \tau^3$, as computed by Rutkevich \cite{RutkevichMesonSpectrumContinuum}.

\section*{References}

\bibliographystyle{iopart-num}
\bibliography{biblio}

\end{document}